\documentclass[%
 reprint,
superscriptaddress,
 amsmath,amssymb,
 aps,
]{revtex4-2}

\usepackage{graphicx}
\usepackage{times}
\usepackage{dcolumn}
\usepackage{bm}

\begin{document}

\preprint{APS/123-QED}

\title{Fundamental measure theory for predicting many-body correlation functions}

\author{Ilian Pihlajamaa}
 \affiliation{Soft Matter and Biological Physics, Department of Applied Physics, Eindhoven University of Technology,
P.O. Box 513, 5600 MB Eindhoven, Netherlands}
\author{Teunike A. van de Pol}
 \affiliation{Soft Matter and Biological Physics, Department of Applied Physics, Eindhoven University of Technology,
P.O. Box 513, 5600 MB Eindhoven, Netherlands}
\author{Liesbeth M.C. Janssen}
 \affiliation{Soft Matter and Biological Physics, Department of Applied Physics, Eindhoven University of Technology,
P.O. Box 513, 5600 MB Eindhoven, Netherlands}
\affiliation{Institute for Complex Molecular Systems, Eindhoven University of Technology, P.O. Box 513, 5600MB Eindhoven, The Netherlands 
}
\date{\today}

\begin{abstract}
We study many-body correlation functions within various Fundamental Measure Theory (FMT) formulations and compare their predictions to Monte Carlo simulations of hard-sphere fluids. FMT accurately captures the qualitative behavior of three- and four-body structure, particularly at low and intermediate wavevectors. At higher wavevectors, the predictions of FMT vary in quantitative accuracy. We show that the dominant contributions to the four-point structure factor arise from direct triplet correlations, allowing the evaluation of four-point correlations to be greatly simplified. In glass-forming liquids at high volume fractions, FMT correctly reproduces deviations from the convolution approximation, highlighting FMT's ability to capture growing structural multipoint correlations upon supercooling. 
\end{abstract}

\maketitle

\section*{Introduction} 

Liquids and glasses lack the long-range order of crystalline solids, yet they exhibit local structure due to the interactions among their constituent particles. Understanding this structure is essential for explaining processes such as freezing, supercooling, and vitrification. One of the most common means to describe the structure is in terms of pair correlation functions \cite{HansenMcDonald2006}. However, this pair-level view is typically insufficient to fully describe very dense systems, where many-body correlations tend to play an important role \cite{Groot1987, Tanaka2013}.
A notable example is the supercooled liquid state, where a growth is observed in static correlation lengths, but only when higher-order correlations are considered \cite{Tanaka2010,Coslovich2011,Leocmach2013}. Indeed, it has been found that triplet correlations significantly influence predictions of the glass transition \cite{Sciortino2001,Ayadim2011,Luo2022}, and four-point correlations show complex long-range ordering in mixtures of liquids \cite{ZhangKob2020}. Despite their importance, many-body correlations remain difficult to extract from experiments and simulations, particularly at long ranges \cite{Bildstein1993,pihlajamaa2023}.

This difficulty motivates the need for accurate theoretical frameworks that can analytically predict the many-body structure of liquids. Classical density functional theory offers such a framework, expressing the free energy of a liquid as a functional of its density profile \cite{Evans1978,HansenMcDonald2006,Mulero2008}. While exact functionals are known in one-dimensional systems \cite{Roth2006}, approximations must be used in higher dimensions. For example, common approaches include the local density approximation, mean-field theory, and weighted density approximation. For the archetypal case of hard spheres, the most successful approach is fundamental measure theory (FMT) \cite{Cuesta2002,Roth2010}.

Fundamental measure theory was first developed by Yaakov Rosenfeld \cite{Rosenfeld1989}, who realized that the
free energy of a hard-sphere liquid could be expressed using the geometric properties (i.e. fundamental measures) of single spheres. FMT has been successfully applied in many situations, such as predicting the density profile of a fluid near a hard wall \cite{Roth2010}, calculating the surface tension of a
liquid, and describing adsorption to substrates \cite{Bernet2020}. However, the original FMT struggles to accurately capture phase transitions such as freezing \cite{Roth2010}. Several refinements have been proposed to address this. Some retain Rosenfeld’s geometric approach but incorporate more accurate equations of state \cite{Roth2002,hansen2006density,malijevsky2006alternative}, while others introduce new mathematical ingredients. 
Most notably,  \citet{Tarazona2000} introduced tensorial weighted densities that can in more detail capture  local constraints in environments of varying dimensions. The scope of FMT has furthermore been broadened beyond hard spheres. In particular, Rosenfeld’s approach has been adapted to account for more complex liquids, such as those made from ‘sticky’ hard spheres or Lennard-Jones particles \cite{HansenGoosWettlaufer2011, Finster2022, Yu2009,Rosenfeld1993,Schmidt1999,Bernet2017}.

The introduction of tensorial weighted densities in FMT suggests that modern formulations may correctly encode local geometric constraints relevant for higher-order structure. Whether this capability yields accurate bulk many-body correlations remains unclear. This is especially relevant for the supercooled state, where the emergence of many-body structure has been determined to be an important precursor and predictor for the dramatic slowdown \cite{Schoenholz2016Softness, Cubuk2015PRL, Royall2008NatMater, RoyallWilliams2015PhysRep,Leocmach2012NatCommun, TongTanaka2018PRX, TongTanaka2019NatCommun, Boattini2020NatCommun, Boattini2021PRL, Jung2023PRL, JackDunleavyRoyall2014PRL, Watanabe2011NatMater}. This gap motivates a direct benchmark of FMT against simulation-derived many-body direct correlation functions. 

In this study, we therefore evaluate the ability of fundamental measure theories to predict many-body direct correlation functions in homogeneous hard-sphere liquids at densities near freezing. We consider Rosenfeld’s original formulation and several of its extensions, and derive analytical expressions for pair, triplet, and four-point direct correlation functions. These are compared against numerical results obtained from simulated structure factors. Finally, we investigate whether FMT can capture the structural signatures of supercooling, which are known to manifest primarily beyond three-body correlations \cite{pihlajamaa2023}.

\section*{Theory}

In fundamental measure theory, the excess free energy functional of a hard-sphere system is expressed as the real-space integral over a local function of weighted densities. These weighted densities are obtained as convolutions of the one-body density $\rho(\mathbf{r})$ with a set of weight functions $\omega_\alpha(\mathbf{r})$ that reflect the geometrical measures of the particle of interest. In general, the excess free energy functional $F_{\text{exc}}[\rho]$ is written as the integral over the excess free energy density $\Phi(\mathbf{r})=\Phi\bigl(\{n_\alpha[\rho](\mathbf{r})\}\bigr)$,

\begin{equation}
F_{\text{exc}}[\rho] = \int \mathrm{d}\mathbf{r} \; \Phi\bigl(\{n_\alpha[\rho](\mathbf{r})\}\bigr),
\end{equation}
where the weighted densities $n_\alpha$ are functionals of $\rho$ and functions of $\mathbf{r}$, given by
\begin{equation}
n_\alpha[\rho](\mathbf{r}) = \int \mathrm{d}\mathbf{r}' \, \rho(\mathbf{r}')\, \omega_\alpha(\mathbf{r} - \mathbf{r}').
\end{equation}
Here $\alpha$ is a label that identifies the particular weight function $\omega_\alpha$.
The $n$-body direct correlation functions $c_n$ are key quantities in density functional theory and FMT, as they describe the response of the system to density perturbations. These functions are derived from the excess free energy functional. In turn, they are related to the $n$-body density through the Ornstein-Zernike relations.

The second-order direct correlation function $c_2(\mathbf{r}, \mathbf{r}')$, often referred to as \textit{the} direct correlation function, is given by the second functional derivative of the excess free energy:
\begin{equation}
c_2(\mathbf{r}, \mathbf{r}') = - \beta\frac{\delta^2 F_{\text{exc}}[\rho]}{\delta \rho(\mathbf{r}) \delta \rho(\mathbf{r}')}.
\end{equation}
Using the FMT expression for $F_{\text{exc}}$, we obtain
\begin{equation}
\begin{split}
  c_2(\mathbf{r}_1, \mathbf{r}_2) =& -\beta\int\mathrm{d}\mathbf{r}\sum_{\alpha, \beta} \frac{\partial^2 \Phi}{\partial n_\alpha \partial n_\beta}(\mathbf{r})\\ &\qquad\times\omega_\alpha(\mathbf{r}_1 - \mathbf{r}) \omega_\beta(\mathbf{r}_2 - \mathbf{r}).  
\end{split}
\end{equation}

Because of the convolutional product, it is convenient to express $c_2$ in Fourier space. Taking the Fourier transform, and assuming that the density field is homogeneous, we find
\begin{equation}
\tilde{c}_2(\mathbf{k}_1, \mathbf{k}_2) = -\beta\delta_{\mathbf{k}_1+\mathbf{k}_2, \mathbf{0}}\sum_{\alpha, \beta} \frac{\partial^2 \Phi}{\partial n_\alpha \partial n_\beta} \tilde{\omega}_\alpha(\mathbf{k}_1) \tilde{\omega}_\beta(\mathbf{k_2}).
\end{equation}
Here, $ \tilde{\omega}_\alpha(\mathbf{k}) $ represents the Fourier transform of the weight function $ \omega_\alpha(\mathbf{r}) $, and the Kronecker delta ensures momentum conservation.

Similar to the two-body case, the $n$-body direct correlation function in Fourier space reads
\begin{equation}\label{eq:cn}
\begin{split}
    \tilde{c}_n(\mathbf{k}_1, \dots, \mathbf{k}_n) &= -\beta\delta_{\mathbf{k}_1+\ldots+\mathbf{k}_n, \mathbf{0}}\\&\times\sum_{\alpha_1, \dots, \alpha_n} \frac{\partial^n \Phi}{\partial n_{\alpha_1} \dots \partial n_{\alpha_n}} \prod_{i=1}^{n} \tilde{\omega}_{\alpha_i}(\mathbf{k}_i).
\end{split}
\end{equation}
Through the many-body Ornstein-Zernike relations, these are related to the many-body structure factors $S_n(\mathbf{k}_1,\ldots,\mathbf{k}_n)$. These can be accessed experimentally by scattering experiments, because they encode correlations between the density field at different wavevectors. Specifically, defining the Fourier transform of the density field $\tilde{\rho}(\mathbf{k})=\sum_{j=1}^N\exp(i \mathbf{k}\cdot \mathbf{r}_j)$, in which $\mathbf{r}_j$ is the position vector of particle $j$, in a system of $N$ particles, the many-body structure factors are defined as
\begin{equation}\label{eq:Sn}
S_n(\mathbf{k}_1,\ldots,\mathbf{k}_n) = \frac{1}{N}\left<\prod_{l=1}^n\tilde{\rho}(\mathbf{k}_l)\right>.
\end{equation}
In the homogeneous single-component bulk phase, the two-, three-, and four-body Ornstein-Zernike relations between $c_n$ and $S_n$ are \cite{coslovich2013static, pihlajamaa2023}
\begin{equation}
    S_2(k) = \frac{1}{1 - \rho c_2(k)},
    \end{equation}
\begin{equation}
     \begin{split}
             S_3(\mathbf{k}_1, \mathbf{k}_2) &= S_2(k_1)S_2(k_2)S_2(|\mathbf{k}_1+\mathbf{k}_2|)\\&\qquad\times(1+\rho^2c_3(\mathbf{k}_1, \mathbf{k}_2)),
     \end{split}   
\end{equation}

\begin{equation}
    \begin{split}
  S_4&(\mathbf{k}_1,\mathbf{k}_2,\mathbf{k}_3) = 
  \\  &\left( 1 + \rho^2c_3(\mathbf{k}_1,\mathbf{k}_2) \right)
   S_2(k_1)S_2(k_2)S_3(\mathbf{k}_1+\mathbf{k}_2, \mathbf{k}_3) \\
 + & \left(1 + \rho^2 c_3(\mathbf{k}_1, \mathbf{k}_3)\right)
   S_2(k_1)S_3(\mathbf{k}_1+\mathbf{k}_3, \mathbf{k}_2)S_2(k_3) \\
 + &\left( 1 + \rho^2c_3(\mathbf{k}_2, \mathbf{k}_3) \right)
   S_3(\mathbf{k}_2+\mathbf{k}_3, \mathbf{k}_1)S_2(k_2)S_2(k_3) \\
  -& \left( 2 - \rho^3 c_4(\mathbf{k}_1,\mathbf{k}_2,\mathbf{k}_3) \right)\\&\quad\times
   S_2(k_1)S_2(|\mathbf{k}_1+\mathbf{k}_2+\mathbf{k}_3|)S_2(k_2)S_2(k_3),
 \end{split}
\end{equation}
where we have explicitly set $\mathbf{k}_n=-\sum_{j=1}^{n-1}\mathbf{k}_j$ and removed it as a function argument to enforce translational invariance in bulk.
These expressions provide the basis for computing three- and four-body correlations and structure factors with FMT. The higher-order structure factors are often approximated purely in terms of $S_2$ by assuming that $c_{n>2} = 0$, an approach known as the convolution approximation \cite{jackson1962energy}. 

Different versions of FMT propose different forms for the free energy density $ \Phi $, using different sets of weight functions $\omega_\alpha$. 
In Rosenfeld's (R's) original formulation \cite{Rosenfeld1989}, the free energy density is given by
\begin{equation}
\begin{split}
    \Phi_{\text{R}} &= -n_0 \ln(1 - n_3) + \frac{n_1 n_2 - \mathbf{n}_{\mathrm{v}1} \cdot \mathbf{n}_{\mathrm{v}2}}{1 - n_3} \\ &\qquad+ \frac{n_2^3 - 3 n_2 \, \mathbf{n}_{\mathrm{v}2} \cdot \mathbf{n}_{\mathrm{v}2}}{24 \pi (1 - n_3)^2},
\end{split}
\end{equation}
where $ n_0 $, $ n_1 $, $ n_2 $, and $ n_3 $ are the scalar weighted densities, and $ \mathbf{n}_{\mathrm{v}1}$ and $\mathbf{n}_{\mathrm{v2}} $ are the vector weighted densities. This formulation recovers the Percus--Yevick (PY) compressibility equation of state in the bulk.

Kierlik and Rosinberg \cite{KierlikRosinberg1990} proposed an alternative formulation that uses only scalar weight functions. 
In this version, the vector contributions are absorbed into redefined scalar weighted densities so that the same bulk thermodynamics is recovered. Because it recovers exactly the correlation functions generated by Rosenfeld's original FMT \cite{phan1993equivalence}, the formulation of Kierlik and Rosinberg is not treated separately here.

Subsequent modifications of Rosenfeld's FMT have been introduced to improve thermodynamic accuracy and consistency. In particular, the White Bear versions modify the free energy density by incorporating the more accurate 
Boublík–Mansoori–Carnahan–Starling–Leland equation of state  \cite{boublik1970hard, mansoori1971equilibrium}.

For the White Bear Mark I (WB1) version, the free energy density is

\begin{equation}
\begin{split}
\Phi_{\text{WBI}} &= -n_0 \ln(1 - n_3) 
+ 
\frac{n_1 n_2 - \mathbf{n}_{\mathrm{v}1} \cdot \mathbf{n}_{\mathrm{v}2}}{1 - n_3} \\
&+ 
\frac{n_2^3 - 3 n_2 \, \mathbf{n}_{\mathrm{v}2} \cdot \mathbf{n}_{\mathrm{v}2}}{36 \pi n_3^2 (1 - n_3)^2}\left(n_3+(1-n_3)^2\ln(1-n_3)\right).
\end{split}
\end{equation}
The White Bear Mark II (WBII) version adds an extra term to reproduce the generalized Carnahan-Starling equation of state, which results in \cite{hansen2006new}
\begin{align}
\Phi_{\mathrm{WBII}} =& -n_0\ln(1-n_3) \nonumber\\\nonumber
&+ \left(1+\frac{n_3^2}{9} \eta^{\mathrm{WBII}}_2(n_3)  \right)\frac{n_1 n_2 - \mathbf{n}_{\mathrm{v}1} \cdot \mathbf{n}_{\mathrm{v}2}}{1-n_3}\\
&+ \left(1- \frac{4n_3}{9} \eta^{\mathrm{WBII}}_3(n_3) \right)\frac{n_2^3 - 3n_2 \, \mathbf{n}_{\mathrm{v}2} \cdot \mathbf{n}_{\mathrm{v}2}}{24\pi(1-n_3)^2}\,,
\end{align}
where
\begin{align}
    \eta^{\mathrm{WBII}}_2(n_3) =& \frac{6n_3 - 3n_3^2 + 6(1-n_3)\ln(1-n_3)}{n_3^3}\,,\\
\eta^{\mathrm{WBII}}_3(n_3) =& \frac{6n_3 - 9n_3^2 + 6n_3^3 + 6(1-n_3)^2\ln(1-n_3)}{4n_3^3}\,.
\end{align}

\citet{Tarazona2000} (T) has enhanced Rosenfeld's scalar formulation by introducing a tensorial weighted density $\mathbf{n}_{\mathrm{m2}}$, by requiring that the functional reduces upon confinement to the exactly known expression in the zero-dimensional case. This has enabled a more accurate description of anisotropic correlations, particularly in crystalline phases of hard spheres. In terms of this new tensorial density and the old scalar and vector densities, the resulting free energy density reads
\begin{equation}
\begin{split}
    &\Phi_{\text{T}} = -n_0 \ln(1 - n_3) + \frac{n_1 n_2 - \mathbf{n}_{\mathrm{v}1} \cdot \mathbf{n}_{\mathrm{v}2}}{1 - n_3} \\
    &+ \frac{n_2^3 - 3 n_2 \, \mathbf{n}_{\mathrm{v}2} \cdot \mathbf{n}_{\mathrm{v}2} + \frac{9}{2}\left[\mathbf{n}^T_{\mathrm{v}2}\mathbf{n}_{\mathrm{m}2}\mathbf{n}_{\mathrm{v}2}- \mathrm{tr}\left({\mathbf{n}_{\mathrm{m}2}^3}\right)\right]}{24 \pi (1 - n_3)^2}.
\end{split}
\end{equation}
The tensorial contributions can also be added to the other free energy densities in order to make use of more accurate equations of state \cite{hansen2006density, oettel2010free}. For example, we denote the tensorial version of the White-Bear II functional as WBIIt. 

Later, \citet{malijevsky2006alternative} (M) proposed a free energy density that reproduces the equation of state of \citet{boublik1986equations}, showing very accurate density profiles near a hard wall. This free energy density reads
\begin{align}
\Phi_{\mathrm{M}} =& -n_0\ln(1-n_3) +\frac{n_1 n_2 - \mathbf{n}_{\mathrm{v}1} \cdot \mathbf{n}_{\mathrm{v}2}}{1 - n_3}\\\nonumber
&+ (n_2^3-3n_2 \mathbf{n}_{\mathrm{v}2} \cdot \mathbf{n}_{\mathrm{v}2})\\\nonumber
&\times \frac{8(1-n_3)^2\ln(1-n_3)+8n_3-15n_3^2/2+2n_3^3}{108\pi n_3^2(1-n_3)^2}\,.
\end{align}

Recently, \citet{lutsko2020explicitly} has generalized Tarazona's expression to enforce global stability, in the sense that the free energy must be bounded from below for any density profile. This leads to a functional form given by
\begin{equation}
\begin{split}
    \Phi_{\text{L}} &= -n_0 \ln(1 - n_3) + \frac{n_1 n_2 - \mathbf{n}_{\mathrm{v}1} \cdot \mathbf{n}_{\mathrm{v}2}}{1 - n_3} \\
    &+ \frac{\frac{8A+2B}{9}n_2^3 - 2 A n_2 \, \mathbf{n}_{\mathrm{v}2} \cdot \mathbf{n}_{\mathrm{v}2} +3A\mathbf{n}^T_{\mathrm{v}2}\mathbf{n}_{\mathrm{m}2}\mathbf{n}_{\mathrm{v}2}}{24 \pi (1 - n_3)^2} \\&+\frac{ -(A+B)n_2\mathrm{tr}\left(\mathbf{n}_{\mathrm{m}2}^2\right)+(2B-A) \mathrm{tr}\left({\mathbf{n}_{\mathrm{m}2}^3}\right)}{24 \pi (1 - n_3)^2},
\end{split}
\end{equation}
which depends on the free parameters $A$ and $B$. Upon the choice  $(A,B)=(3/2, -3/2)$, $\Phi_{\text{L}}$ reduces to Tarazona's expression. Lutsko has shown that this functional is only globally stable if $A,B\geq0$, and proposes the choice $(A,B)=(1,0)$. In a follow-up work \citet{gul2024using} suggests an alternative choice $(A,B) = (1.3, -1.0)$ based on the best incorporation of thermodynamical sum rules. The free energies obtained after substitution of the parameter choices are respectively denoted by (L) and (G).

All of the above free energy densities are written as a function of weighted densities $n_\alpha[\rho](\mathbf{r})$. The associated weights $\omega_\alpha(\mathbf{r})$ are related to the so-called fundamental measures of the sphere. Specifically, their definitions and corresponding Fourier transforms are given by
\begin{align*}
\omega_0(\mathbf{r}) &= \frac{\delta(r-R)}{4\pi R^2}, & \tilde{\omega}_0(\mathbf{k})&=j_0(kR),\\
\omega_1(\mathbf{r}) &= \frac{\delta(r-R)}{4\pi R}, & \tilde{\omega}_1(\mathbf{k})&=Rj_0(kR),\\
\omega_2(\mathbf{r}) &= \delta(r-R), & \tilde{\omega}_2(\mathbf{k})&=4\pi R^2j_0(kR),\\
\omega_3(\mathbf{r}) &= \Theta(R-r),& \tilde{\omega}_3(\mathbf{k})&=\frac{4\pi R^2}{k}j_1(kR),\\
\boldsymbol{\omega}_{\mathrm{v}1}(\mathbf{r}) &= \frac{\mathbf{r}}{r}\frac{\delta(r-R)}{4\pi R}, & \tilde{\boldsymbol{\omega}}_{\mathrm{v}1}(\mathbf{k})&=-R\frac{i\mathbf{k}}{k}j_1(kR),\\
\boldsymbol{\omega}_{\mathrm{v}2}(\mathbf{r}) &= \frac{\mathbf{r}}{r}\delta(r-R), & \tilde{\boldsymbol{\omega}}_{\mathrm{v}2}(\mathbf{k})&=-4\pi R^2\frac{i\mathbf{k}}{k}j_1(kR),
\end{align*}
\vspace{-0.7cm}
\begin{multline}
        \boldsymbol{\omega}_{\mathrm{m}2}(\mathbf{r}) = \left(\frac{\mathbf{r}\mathbf{r}}{r^2}-\frac{\mathbf{I}}{3}\right)\delta(r-R), \\  \tilde{\boldsymbol{\omega}}_{\mathrm{m}2}(\mathbf{k})=-4\pi R^2\left(\frac{\mathbf{k}\mathbf{k}}{k^2}-\frac{\mathbf{I}}{3}\right)j_2(kR), 
\end{multline}
where $j_0(x)=\sin(x)/x$, $j_1(x) = (\sin(x)-x\cos(x))/x^2$, and $j_2(x)=(3\sin(x)-3x\cos(x)-x^2\sin(x))/x^3$ are the first three spherical Bessel functions of the first kind, $\mathbf{r}\mathbf{r}$ and $\mathbf{k}\mathbf{k}$ are dyadic products, and $\mathbf{I}$ is the unit tensor. We note that by opting to work with the traceless tensorial weight $\omega_{\mathrm{m}2}$, the forms of the tensorial functionals above are slightly different than if they had been written in terms of $\omega_{\mathrm{m}2}'= \mathbf{rr} \delta(r-R)/r^2$, which is also used in the literature. Additionally, for the vector- and tensor-like weight functions, not only the sign but also the direction of the wave vector is important for the correctness of the result.

The Fourier representations of the weight functions are now used to construct the direct correlation functions in Fourier space using Eq.~\eqref{eq:cn}, which can then be used to predict the $n$-body structure factors. This procedure yields fully explicit symbolic expressions, and while they can become very long, their evaluation is essentially free in a computational sense.

\section*{Results}
To assess the accuracy of various FMT formulations in capturing $n$-body structural correlations, we compute the direct correlation functions $c_n(k)$ and static structure factors $S_n(k)$ for a hard-sphere fluid in which the particles have diameter $\sigma=1$, which serves as our unit of length. We consider a number density $\rho=0.94$ just below freezing, corresponding to a volume fraction $\eta = \pi\rho/6 \approx 0.49$. For reference, we compare each of the theories to correlation functions measured from standard Metropolis Monte Carlo (MC) simulations of $N=30,000$ particles in the canonical ensemble (NVT). To get sufficient statistics, we perform 100 independent equilibration and production runs of $10^6$ MC sweeps each, and store a snapshot every $10^3$ sweeps during production. Each independent simulation took around 10 CPU hours. From the $10^5$ total snapshots, we compute the structure factors $S_n$ directly from their definition, Eq.~\eqref{eq:Sn} (see Ref.~\cite{pihlajamaa2023} and its supplementary information for more details). In all figures pertaining to the hard-sphere MC simulations, the error bars indicate the standard deviation of the mean across the 100 independent runs. The simulation, analysis, and plotting code can be found in the repository of Ref.~\cite{Pihlajamaa2025FundamentalMeasureTheory}. We start with a comparison for $n=2$, and consider $n=3,\ 4$ later in this section.

\begin{figure}[t]
    \centering
    \includegraphics[width=\linewidth]{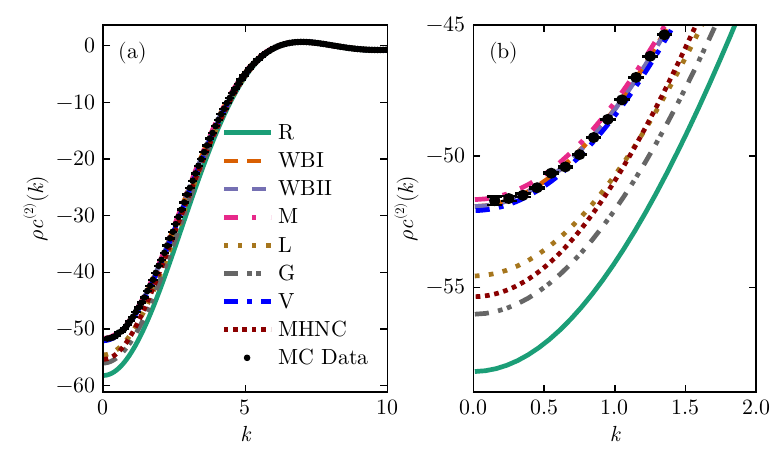}
    \caption{(a) Fourier-space two-body direct correlation function $\rho c_2(k)$ for a hard-sphere fluid at volume fraction $\eta = 0.49$, comparing various theoretical models to Monte Carlo (MC) simulation data (black symbols). (b) Magnification of the low-$k$ region. FMT-based predictions are shown for the original Rosenfeld (R), White Bear I (WBI), White Bear II (WBII),  Malijevský (M) Lutsko (L) and G\"{u}l (G) functionals, alongside results obtained from Ornstein–Zernike closures, specifically the Verlet (V) and Modified Hypernetted Chain (MHNC) theories. For this and all following figures, the error bars denote the standard deviation of the mean from 100 different simulation runs. Here, the error bars are typically smaller than the markers.}
    \label{fig:c2}
\end{figure}

To place the performance of the different FMT formulations for the 2-point bulk structure in theoretical context, we compare the FMT predictions here additionally to the very successful Verlet (V) \cite{verlet1980integral, verlet1981integral} and Modified Hypernetted Chain (MHNC) closures of the Ornstein-Zernike relation \cite{rosenfeld1979theory}, which are go-to methods for predicting pair structure in bulk. The MHNC closure uses a parametrization of the bridge function of hard spheres measured from Monte Carlo (MC) simulations and is therefore expected to offer virtually exact predictions for $r>1$ \cite{malijevsky1987bridge, pihlajamaa2024comparison}. The integral equation theories are solved by the software introduced in Ref.~\cite{pihlajamaa2024comparison}.

\begin{figure}[t]
    \centering
    \includegraphics[width=\linewidth]{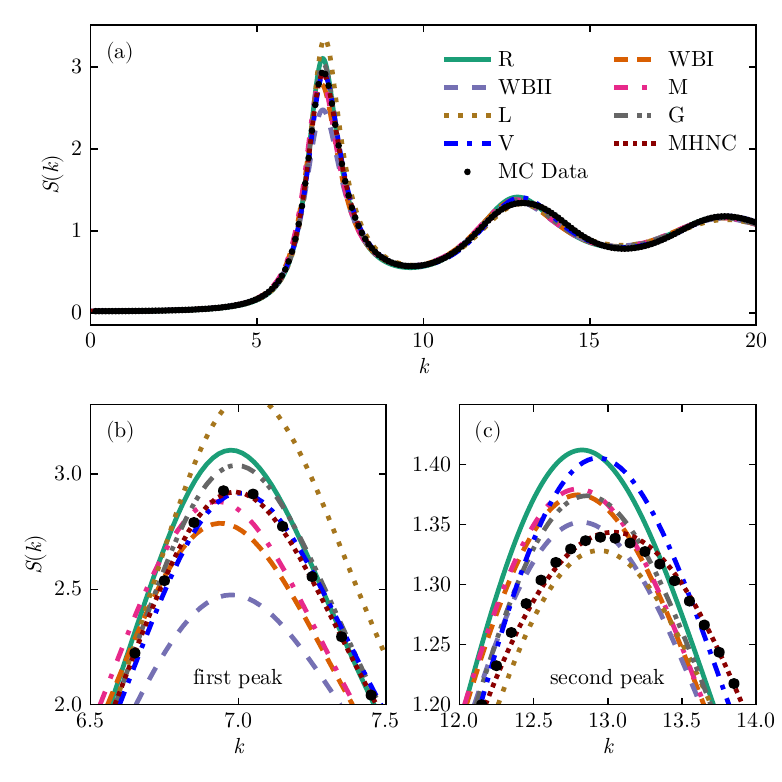}
    \caption{(a) Static structure factor $S_2(k)$ of the hard-sphere fluid at $\eta = 0.49$ as a function of the wavenumber $k$, comparing Monte Carlo data (black circles) to the same theoretical models as in Fig. \ref{fig:c2}. Panels (b) and (c) magnify the first and second peak regions, respectively.}
    \label{fig:S2}
\end{figure}

Figure \ref{fig:c2} shows the two‐body direct correlation function $\rho c_2(k)$ for a hard‐sphere fluid at volume fraction $\eta = 0.49$ comparing Monte Carlo data to a range of theoretical models: Rosenfeld’s original fundamental‐measure theory (R), White Bear I (WBI), White Bear II (WBII), Malijevský’s variant (M), the newer functionals of Lutsko  (L) and G\"{u}l (G), and the two integral‐equation closures. The tensorial theories (T, WBIIt) are not considered here because their tensorial contributions have no effect at the pair level, reducing their predictions to those of their non-tensorial equivalents (R, WBII). Qualitatively, all FMTs show the correct behavior for the direct correlation function. Quantitatively, for the longest wavelengths ($k\to0$), Rosenfeld’s FMT predicts a considerably more negative $c_2(k)$ than the others. This highlights its commonality with Percus–Yevick theory, resulting in an isothermal compressibility that is lower than that computed from MC data \cite{HansenMcDonald2006}. The FMTs of Lutsko and Gül perform better, but still show significant errors in the low-$k$ regime.  In contrast, the White Bear I/II functionals, which enforce the Boublík–Mansoori–Carnahan–Starling–Leland or generalized Carnahan–Starling equation of state \cite{boublik1970hard, mansoori1971equilibrium, hansen2006new}, bring $c_2(0)$ into near‐perfect agreement with simulation. Malijevský’s free energy density, also built on a high‐accuracy bulk equation of state, gives essentially the same low‐$k$ result as WBII, very slightly underestimating $c_2(0)$ compared to WBI/II and the MC data. The Verlet closure, likewise, matches the MC compressibility almost exactly, whereas the MHNC closure fails to accurately capture the small-$k$ behavior because its parametrization was performed for $r>1$ only.
At high $k$ ($k\gtrsim6$), every model’s $c_2(k)$ approaches zero in lockstep with MC. To visualize the differences between the theories, we consider the static structure factor $S(k)=1/(1-\rho c_2(k))$, which amplifies the small differences in $\rho c(k)$ at larger $k$.

Figure \ref{fig:S2} presents the static structure factor $S_2(k)$. The principal diffraction peak at $k\approx7$ attains $S_\text{max}\approx2.9$ in simulation. Rosenfeld, Lutsko, and Gül overestimate this peak ($S_\text{max}\approx3.1$, $S_\text{max}\approx3.5$, and , $S_\text{max}\approx3.0$, resp.), while WBI and WBII underestimate it  ($S_\text{max}\approx2.7$ and $S_\text{max}\approx2.4$, resp.). Malijevský’s form produces much better agreement with simulation data. The Verlet closure matches the first peak with virtually no error, and so does MHNC.

\begin{figure*}[t]
\centering
\includegraphics[width=0.99\linewidth]{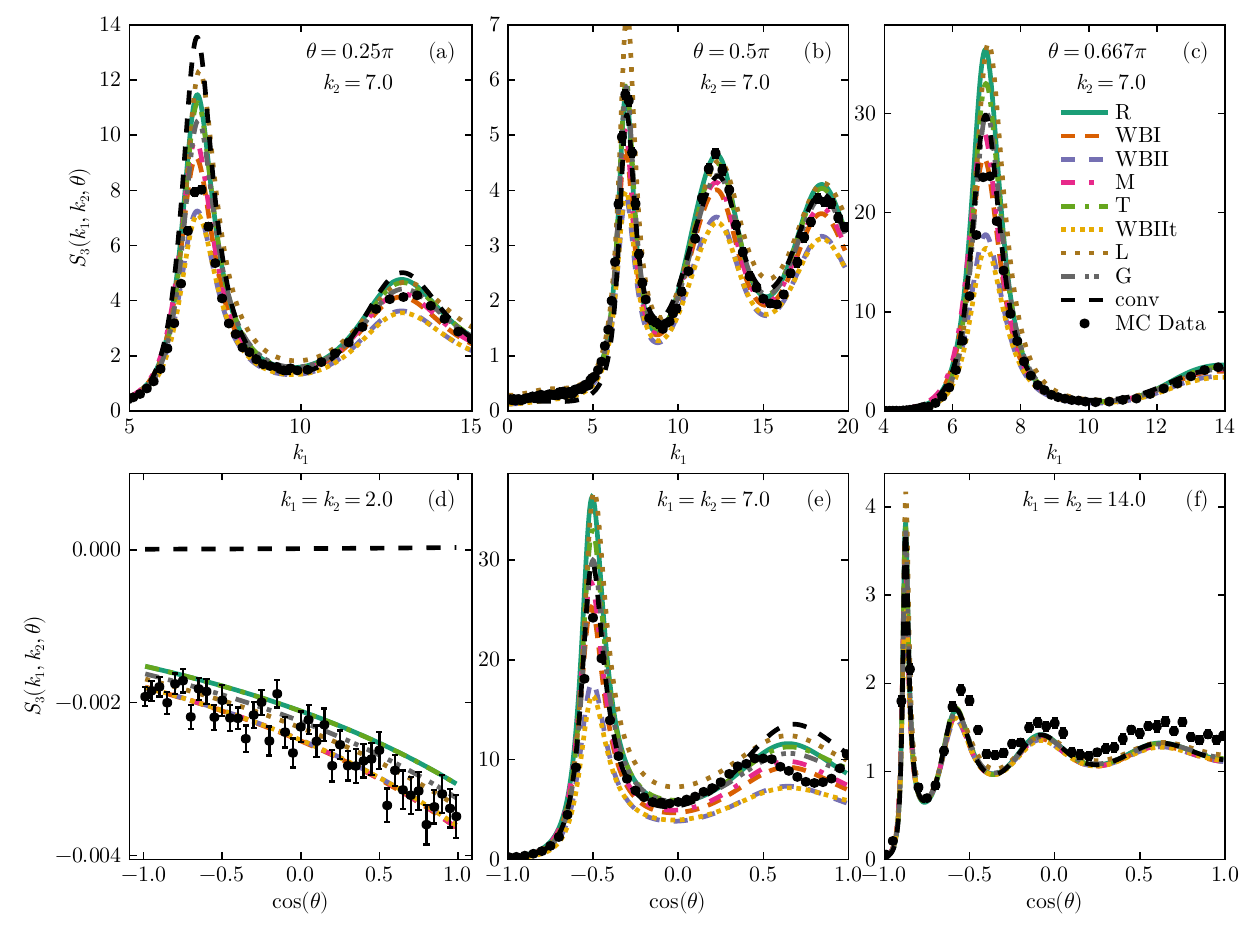}
\caption{Triplet structure factor $S_3(k_1,k_2,\theta)$ at $\eta = 0.49$ from Monte Carlo (MC) simulations (black circles) compared to predictions from the different FMTs. The convolution approximation ($c_3=0$) is also shown, for which we have used the Rosenfeld prediction of $S_2$. Top row: $S_3$ as a function of $k_1$ at fixed $\theta$ and $k_2$. Bottom row: $S_3$ as a function of $\cos\theta$ at fixed $k_1 = k_2$. }
\label{fig:S3}
\end{figure*}

The second peak at $k\approx13$ is subtler ($S_\text{2nd max}\approx1.34$ from MC). Rosenfeld again overshoots ($S_\text{2nd max}\approx1.42$), as do the Verlet closure, Malijevský's functional, WBI, G, and WBII, in order of increasing accuracy. Lutsko's FMT slightly undershoots, and the MHNC relation reproduces the MC data with very high accuracy, which is perhaps not surprising since it incorporates a high-fidelity MC correlation data parameterization. Beyond the second peak, all curves converge toward $S_2(k)\to1$, with decreasing model-to-model variation.

These findings echo the earlier observations that Rosenfeld’s FMT in bulk inherits the inferior accuracy of Percus-Yevick theory and overestimates correlations in the fluid, and that White Bear variants improve thermodynamic consistency but can underestimate the pair correlations \cite{Roth2010}. Overall, while the more complex functionals reproduce the $c_2(k\to0)$ behavior better than the original theory by Rosenfeld, their improved accuracy is not universally retained when considering shorter wavelengths.

\subsection*{Triplet correlations}

We now turn to the analysis of triplet correlations in the bulk hard-sphere fluid. Figure~\ref{fig:S3} shows the triplet structure factor $S_3$ as a function of $k_1$ for fixed $k_2$ and angle $\theta$ (a--c), and in the isosceles triangle configuration, i.e.~as a function of $\cos\theta$ for fixed $k_1 = k_2$ (d--f). The predictions from different FMT functionals are compared to Monte Carlo (MC) data, which are consistent with earlier data from the literature \cite{curtin1990triplet, rosenfeld1990free}. All functionals capture the general features of $S_3$ across a wide range of geometries, but differ in the degree to which they reproduce fine structural details.

At low wavenumbers, all FMT formulations provide excellent agreement with MC data. The agreement becomes less precise as $k$ increases. For example, in panels (b--c) and (e--f), where $k_1$ spans intermediate to high values ($k_1 \approx 7 - 14$), the relative accuracy of the FMT predictions varies drastically with the angle and wavenumbers, showing no clear `best' theory. Interestingly, in the small-angle isosceles configuration with wavenumbers around the main diffraction peak $k_1=k_2=7.0$, $\cos\theta\approx1$, all theories show qualitative disagreement with the MC data. We surmise that this is related to earlier observations that the FMTs do not strictly enforce zero probability upon particle overlap \cite{Roth2002, tschopp2021fundamental}.  See also the work of \citet{coslovich2013static} for similar observations.

\begin{figure*}[ht]
\centering
\includegraphics[width=0.99\linewidth]{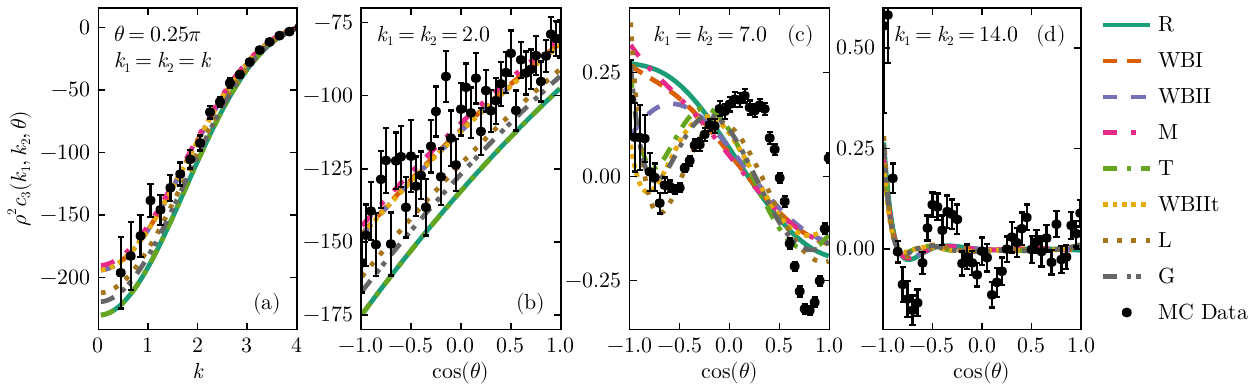}
\caption{Triplet direct correlation function $\rho^2 c_3(k_1,k_2,\theta)$ for a hard-sphere fluid at $\eta = 0.49$. Black symbols show Monte Carlo (MC) results. (a) Direct triplet correlations $c_3$ against $k$ for $k_1 = k_2 = k$ and $\theta = 0.25\pi$. (b–d) $c_3$ as a function of $\cos\theta$ for various $k_1 = k_2$ values. }
\label{fig:C3}
\end{figure*}

Figure~\ref{fig:C3} presents the corresponding triplet direct correlation function $c_3(k_1,k_2,\theta)$, plotted as a function of either $k_1=k_2=k$ or $\cos\theta$. 
At low wavenumbers (panels a, b), all FMT functionals describe the MC data well, with the best predictions given by the WBI/II(t) and M theories. This superiority of WBI/I(t) and M is expected, as the low-$k$ regime is primarily governed by thermodynamic consistency and bulk equation-of-state constraints, which these theories incorporate more accurately.

In contrast, at values of $k$ around the main diffraction peak (panel c), discrepancies between the models become more pronounced. In the isosceles configuration ($k=k_1 = k_2 = 7.0$), the predicted $c_3$ curves differ in both amplitude and angular dependence from the MC data. While all functionals capture the overall magnitude of $c_3$, only the functionals that incorporate tensorial contributions (T, WBIIt, L, and G) show reasonable qualitative agreement with the data.  This mirrors the findings from our pair level analysis above, where we have shown that improved thermodynamic functionals give more accurate $c_2(k)$ at small $k$ but not necessarily so at larger $k$. If we increase $k$ to a value of 14.0, all FMTs predict that direct triplet correlations are only relevant for $\cos(\theta)\approx-1$,  which is not in agreement with MC data, which show marked oscillations with $\theta$.  

Overall, all tested FMT formulations capture the qualitative behavior of three-body correlations in the hard-sphere fluid. At long wavelengths (small $k$), the agreement between FMT and simulation is excellent. At higher $k$, performance varies depending on the geometry and specific formulation used. The more recent FMTs, particularly those incorporating tensorial corrections, offer improvements in reproducing the structure and angular dependence of $S_3$ and $c_3$ compared to earlier formulations. 

\subsection*{Four-point correlations}

\begin{figure}[ht]
\centering
\includegraphics[width=0.99\linewidth]{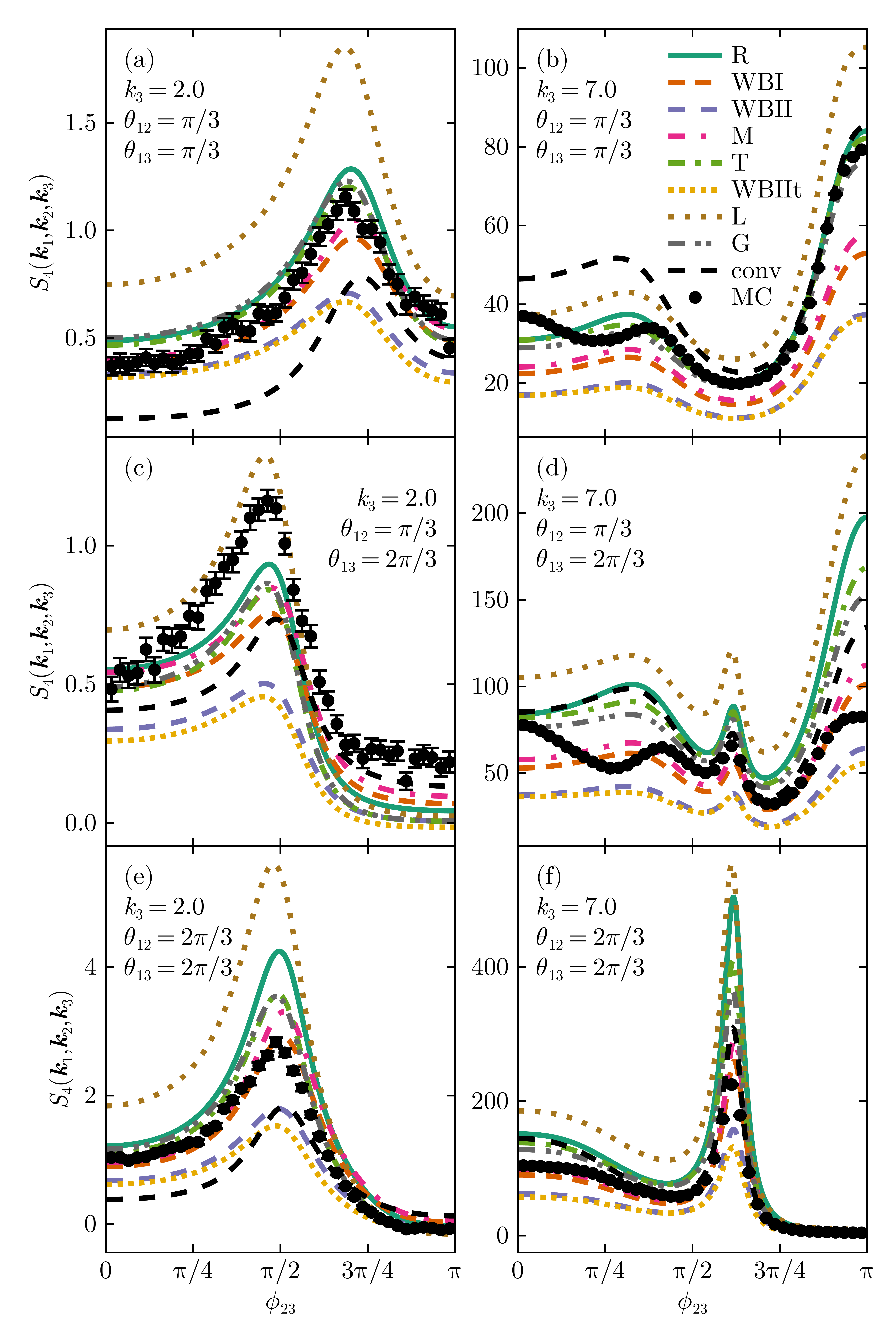}
\caption{Four-body static structure factor $S_4(k_1,k_2,k_3,\theta_{12},\theta_{13},\phi_{23})$ for a hard-sphere fluid at $\eta = 0.49$. Monte Carlo data (black symbols) are compared to various FMT predictions and the convolution approximation obtained with the Rosenfeld $S_2$ (black, dashed). The wavenumbers are fixed at $k_1 = k_2 = 7.0$, while $k_3 \in \{2.0, 7.0\}$ as labeled. Each panel corresponds to a different angular configuration.}
\label{fig:S4_full}
\end{figure}

To assess the predictive power of FMT in capturing higher-order correlations, we now turn to the four-point static structure factor $S_4(k_1,k_2,k_3,\theta_{12},\theta_{13},\phi_{23})$. Figure~\ref{fig:S4_full} shows comparisons between Monte Carlo (MC) simulations and FMT predictions for six different angular configurations, with $k_1 = k_2 = 7.0$, and varying $k_3 \in \{2.0, 7.0\}$. The angles $\theta_{12}$ and $\theta_{13}$ are chosen from $\pi/3$ and $2\pi/3$ because these configurations yielded the largest observed correlations. We expect this is due to the appearance of these angles in hexagonal (FCC/HCP), tetrahedral, and icosahedral motifs prevalent in hard-sphere fluids \cite{leocmach2012roles, berryman2016early,  marin2020tetrahedrality, carter2018structural}. Apart from the six wave vector configurations shown here, we have tested another 92 different sets of angles and wave numbers. The corresponding figures are available at Ref.~\cite{Pihlajamaa2025FundamentalMeasureTheory}. 

Across all panels, FMT successfully captures the general shape and location of the principal features of $S_4$ with minor exceptions. However, as with the triplet correlations, different functionals vary significantly in their quantitative accuracy. The original Rosenfeld (R) formulation and Lutsko's FMT tend to overpredict the peak heights in nearly all cases, while White Bear II and its tensorial version consistently underpredict the correlations. Overall (also considering many other configurations not shown here), we have found that the White Bear I functional gives the best results out of all the functionals for the four-body structure. We believe, however, that this is mostly fortuitous.

Interestingly, the convolution approximation (black dashed lines), which neglects both $c_3$ and $c_4$, performs surprisingly well in reproducing qualitatively the overall angular structure at this density, as was noted before \cite{pihlajamaa2023}. In particular, it gives essentially the same qualitative predictions as the different FMTs. This raises the question of which many-body direct correlation functions are essential for an accurate description of $S_4$.

\begin{figure}[ht]
\centering
\includegraphics[width=0.99\linewidth]{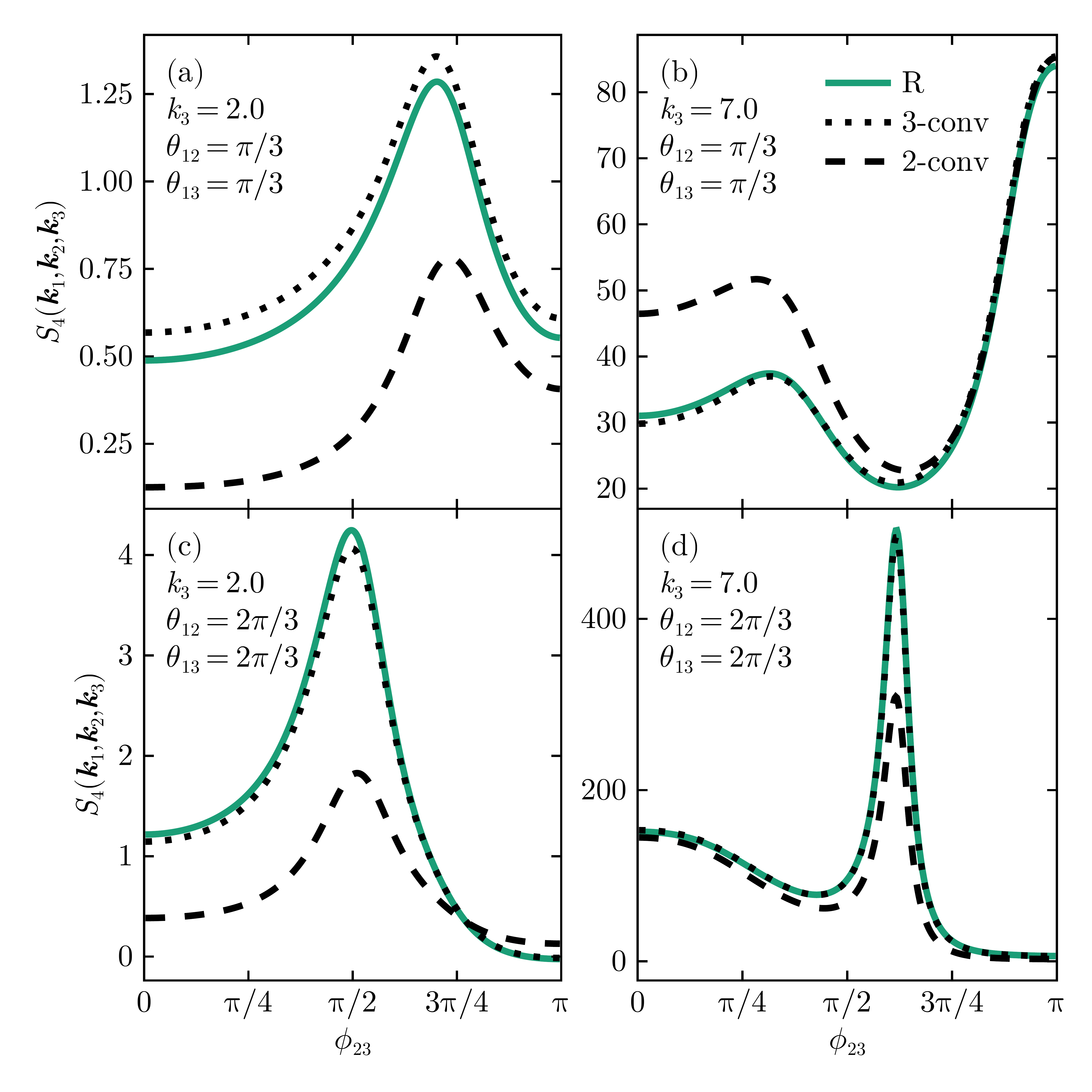}
\caption{Effect of higher-order direct correlation functions on the four-point structure factor $S_4$. Each panel shows $S_4$ computed from the Rosenfeld functional using the full four-body FMT expression (solid lines), with $c_3$ but not $c_4$ retained (dotted, 3-conv.), and with both $c_3 = 0$ and $c_4 = 0$ (dashed, 2-conv.). The wavenumbers are fixed at $k_1 = k_2 = 7.0$, while $k_3 \in \{2.0, 7.0\}$ as labeled.}
\label{fig:S4_conv}
\end{figure}

To investigate the individual contributions of $c_3$ and $c_4$ to the full four-body structure factor, we perform a systematic comparison in Figure~\ref{fig:S4_conv}, where we compute $S_4$ using either full FMT input, only the triplet correlation $c_3$, or a pure convolution approximation with both $c_3 = 0$ and $c_4 = 0$. These calculations are performed using the original Rosenfeld functional. We have verified that the other functionals give qualitatively the same results.

Remarkably, the results show that the correction to the convolution approximation arises almost entirely from the inclusion of $c_3$. The dashed curves (no $c_3$ or $c_4$) consistently misrepresent the magnitude of $S_4$ compared to the full theory, while the dotted curves (only $c_4 = 0$) reproduce the full FMT predictions (green solid curves) almost exactly. This holds across all tested configurations, and is especially pronounced in panels (c--d), where the omission of $c_3$ leads to a dramatic underestimation of the main peak, while the omission of $c_4$ leaves the correlations virtually untouched. These results suggest that in the bulk hard-sphere fluid, four-body correlations are dominated by three-body contributions, and that including $c_4$ yields only marginal improvements.

This finding can be useful for drastically simplifying theoretical treatments of many-body correlations in fluids. The inclusion of $c_3$ already suffices to capture the majority of the structure in $S_4$, substantially reducing the complexity of the required calculations. Given that the full expression for $c_4$ is cumbersome to evaluate and introduces only minimal corrections, neglecting $c_4$ should be a justifiable approximation for simple fluids.

\subsection*{Four-point correlations in the supercooled regime}

\begin{figure*}[ht]
\centering
\includegraphics[width=0.999\linewidth]{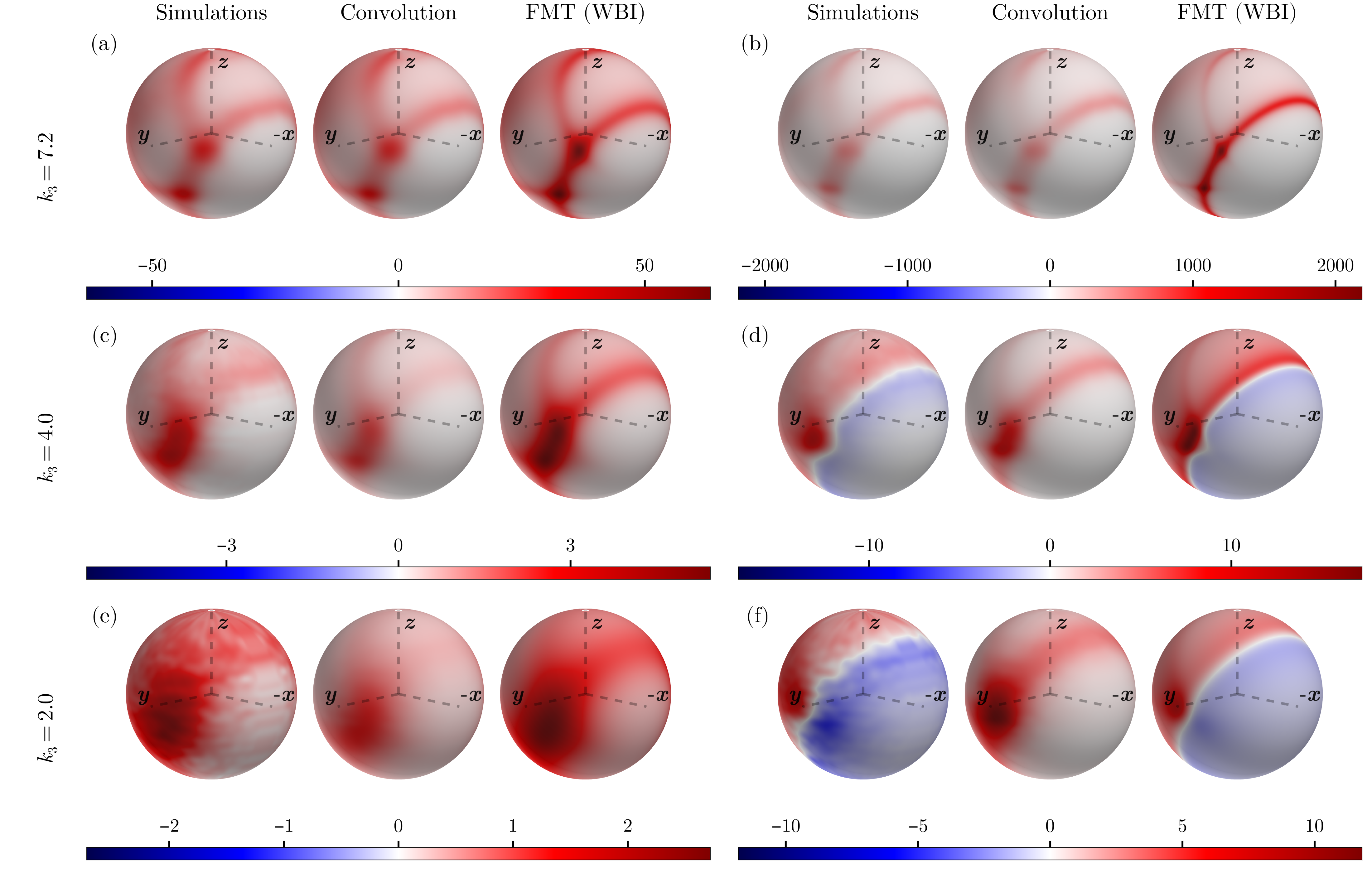}
\caption{Four-point static structure factor $S_4(k_1,k_2,k_3,\theta_{12},\theta_{13},\phi_{23})$ shown on the surface of a unit sphere parameterized by $\theta_{13}$ and $\phi_{23}$ for $k_1 = k_2 = 7.2$ and $\theta_{12} = 2\pi/3$. Rows correspond to different values of $k_3 \in \{7.2, 4.0, 2.0\}$. Columns compare Monte Carlo simulation (left), convolution approximation (center), and FMT predictions using the White Bear I functional (right). Left panels (a, c, e): $\eta = 0.45$; right panels (b, d, f): $\eta = 0.58$. The simulation data are reused from Ref.~\cite{pihlajamaa2023}.}
\label{fig:S4supercooled}
\end{figure*}

To further test the performance of FMT, we consider supercooled liquids, extending our analysis to volume fractions beyond the freezing point. To this end, we simulate a 6\% polydisperse quasi-hard-sphere system, as introduced in Ref.~\cite{weysser2010}, which avoids crystallization and allows us to probe the structure of a metastable, amorphous fluid up to volume fractions of $\eta = 0.58$. In this system, the $N=1000$ particles interact with an inverse power-law potential $u(r)\propto r^{-36}$.  At this high density, corresponding to the supercooled regime, growing structural correlations are known to emerge \cite{Tanaka2010, tanaka2012, BerthierKob2012, LeocmachTanaka2012, Biroli2013, Li2014, RussoTanaka2015, Hallett2018, Tah2018}, and four-point structure factors provide a natural way to probe such effects \cite{pihlajamaa2023}.

We focus on the four-point static structure factor $S_4(k_1, k_2, k_3, \theta_{12}, \theta_{13}, \phi_{23})$, shown in Fig.~\ref{fig:S4supercooled}, plotted on the surface of a unit sphere defined by polar angle $\theta_{13}$ and azimuthal angle $\phi_{23}$ for fixed wavenumbers $k_1 = k_2 = 7.0$ and polar angle $\theta_{12} = 2\pi/3$. Each row corresponds to a different value of $k_3$, ranging from 7.2 (top) to 2.0 (bottom), and we compare three different data sets: MC simulations (left), the convolution approximation obtained with the simulation $S_2$ (center), and predictions from the White Bear I FMT functional (right). The left column (panels a, c, e) shows results at a moderate volume fraction of $\eta = 0.45$, while the right column (panels b, d, f) shows the corresponding results at a supercooled state point $\eta = 0.58$.

At moderate density, both the convolution approximation and FMT capture the general features of $S_4$ observed in simulation, with quantitative differences remaining modest. However, at $\eta = 0.58$, the structure of $S_4$ changes significantly. In particular, the simulations show a clear qualitative change in correlations at low and intermediate wavevectors ($k_3 = 2.0, 4.0$), and the convolution approximation fails to reproduce this behavior. These deviations from the convolution approximation have been reported recently \cite{pihlajamaa2023}, where they were attributed to the emergence of genuine many-body correlations that cannot be constructed from products of pair correlations alone.

Remarkably, the FMT-based predictions are able to reproduce many of the qualitative trends seen in the simulations. In all panels, the WBI functional captures the qualitative structure of $S_4$ more accurately than the convolution approximation, particularly at $\eta = 0.58$ and low $k_3$. This suggests that fundamental measure theory, while developed primarily for equilibrium fluids, can also provide useful structural predictions in the supercooled regime. The qualitative accuracy at low $k$ is particularly noteworthy, as this is precisely the regime where the convolution approximation was previously found to break down most strongly \cite{pihlajamaa2023}.

The ability of FMT to qualitatively improve upon the convolution approximation is independent of which FMT functional is chosen, as we find that each FMT considered in this work correctly reproduces the low-$\mathbf{k}_3$ anisotropy of $S_4$. Quantitatively, the FMTs deviate in the magnitude of their predictions of $S_4$. Because the simulation data used here are not of single-component hard spheres but of polydisperse nearly-hard spheres, we have restricted our comparison with single-component FMTs to the qualitative level. A proper quantitative comparison between the different FMTs would require using their generalizations to multi-component systems. We expect that our use of the single-component theories for application in mixtures also causes the large overestimation of the correlations shown in panels \ref{fig:S4supercooled}(a--d). 

These findings provide additional evidence for the (modestly) growing static length scales found in deeply supercooled liquids \cite{Tanaka2010, Coslovich2011, BerthierKob2012, LeocmachTanaka2012, Biroli2013, Li2014, RussoTanaka2015, Hallett2018, Tah2018}. The ability of FMT to reproduce these features implies that it captures essential geometric constraints and correlations, even beyond equilibrium. It remains an open challenge to systematically derive the connection between such growing four-point structure and dynamic slowdowns \cite{Schoenholz2016Softness, Cubuk2015PRL, Royall2008NatMater, RoyallWilliams2015PhysRep,Leocmach2012NatCommun, TongTanaka2018PRX, TongTanaka2019NatCommun, Boattini2020NatCommun, Boattini2021PRL, Jung2023PRL, JackDunleavyRoyall2014PRL, Watanabe2011NatMater, janssen2018mode}, but our results suggest that higher-order direct correlation functions could play a central role in such a theory. 

\section*{Conclusion}

In this work, we have developed and tested analytical expressions for many-body direct correlation functions up to fourth order, within several formulations of fundamental measure theory, and evaluated their capacity to reproduce structural correlations in dense hard-sphere fluids. By comparing FMT predictions to Monte Carlo simulation data for two-, three-, and four-body structure factors, we have shown that FMT captures a substantial portion of the many-body physics, especially at low and intermediate wavevectors. Our results confirm that FMT functionals, particularly those incorporating tensorial contributions and accurate equations of state, can reproduce key features of triplet and even four-point correlations in the bulk. Their performance generally is best at low wavenumbers $k$, where thermodynamics dominate the behavior, and starts to decrease at larger $k$, where the theory loses accuracy due to more nuanced structural changes.

Importantly, we have demonstrated that for predicting the four-point static structure factor $S_4$ in equilibrium liquids, the inclusion of the triplet direct correlation function $c_3$ alone is sufficient to account for the dominant features, with $c_4$ contributing only marginal corrections. This insight significantly simplifies theoretical modeling of higher-order structure in dense fluids.

Extending our analysis into the supercooled regime, we have found that FMT can qualitatively reproduce the growing structural correlations that emerge in this metastable state. In particular, FMT captures qualitative deviations from the convolution approximation at low wavevectors, previously identified as a signature of many-body amorphous ordering \cite{ZhangKob2020, pihlajamaa2023}. These findings suggest that FMT may serve not only as an accurate equilibrium theory, but also as a valuable tool for studying the structural origin of glassy dynamics.

\section*{Acknowledgments} 
We would like to thank Martin Oettel for useful discussions and Corentin Laudicina for his valuable insights and careful reading of the manuscript. Additionally, we acknowledge the Dutch Research Council (NWO) for financial support through a Vidi grant (IP and LMCJ). 

\section*{Data availability}
The data that support the findings of this article and the code used to generate them are openly available \cite{Pihlajamaa2025FundamentalMeasureTheory}.

\bibliography{apssamp}

\end{document}